      [1999/12/01 v1.4c Il Nuovo Cimento]

\documentclass[varenna]{cimento}
\usepackage{graphicx}

\title{A survey for Ly$\alpha$ galaxies at z $\sim$ 3.1}

\author{G. Ventimiglia, M. Arnaboldi}

\institute{ESO \\ Karl-Schwarzschild-Strasse 2\\ D-85748 Garching bei Muenchen}

\author{O. Gerhard}

\institute{Max Planck Institut fuer Extraterrestrische Physik\\
              Giesenbachstrasse\\ D-85741 Garching bei Muenchen}

\shortauthor{G. Ventimiglia, M. Arnaboldi \atque O. Gerhard}

\begin{document}

\maketitle

 \begin{abstract}
   We describe the results of a deep survey for Ly$\alpha$ emission
   line galaxies at z $\sim$ 3.1, carried out with the multislit
   imaging spectroscopy (MSIS) technique, with the FORS2 spectrograph
   on VLT-UT1.  We discuss the criteria used to select the emission
   line galaxies and present the main physical characteristics of the
   sample: redshift, observed flux and equivalent width distributions.
 \end{abstract}

\section{Introduction}
In the last decades our knowledge of the high redshift (z $>$ 2)
universe has significantly increased. The observational technique that
allowed such galaxies to be found, in a significant number, is the
dropout technique \cite{ref:Giavalisco02a}. It detects Lyman Break
galaxies by the flux discontinuity due to their Lyman limit absorption
\cite{ref:Steidel96a,ref:Steidel96b}. Since 1998, narrow band surveys
reported the detection of Ly$\alpha$ emission from objects in
the redshift range 2.4 $<$ z $<$ 6.5
\cite{ref:Hu98,ref:Ciardullo02,ref:Ajiki03, ref:Ouchi05,ref:Ouchi08,
  ref:Gronwall07,ref:Schaerer07}. The Ly$\alpha$ emission not only allows
galaxies to be detected at very high redshifts, but also gives a valuable
star formation diagnostic and facilitates the study of large scale
structures at high redshift. With spectroscopic surveys the
Ly$\alpha$ emission line profiles can be studied \cite{ref:Kudritzki00,
  ref:Shapley03,ref:Martin08,ref:Rauch08}. This, in turn, provides the
possibility of testing models of the physical parameters of the Lyman
$\alpha$ emitters and to derive constraints on their stellar
populations and their gas and dust content \cite{ref:Verhamme08}.

Here we present the results of a survey at z $\sim$ 3.1 carried out
with the multislit imaging spectroscopy technique \cite{ref:Gerhard05}
(MSIS) with FORS2 on UT1.

\section{Observational set up}
\subsection{The Multislit Imaging Spectroscopy Technique}
MSIS is a blind search technique that consists of the combined use of
a mask of parallel slits, a narrowband filter, and a dispersing
element. It obtains the spectra of all emission line objects that
happen to lie behind the mask slits. Our main purpose is to detect the
[OIII] $\lambda$5007 line emission for Planetary Nebulae (PNs), in
order to study the kinematic properties of the Intracluster light
(ICL) in the central regions of nearby ($<$ 100 Mpc) clusters of
galaxies \cite{ref:Gerhard07,ref:Ventimiglia08}. By dispersing the sky
noise on many pixels, the technique enables measurements of very faint
fluxes. Thus MSIS surveys are also suitable for the detection of the
redshifted $1216$ \AA\ emission line from high-redshift Ly$\alpha$
galaxies.

\subsection{Observational set up}
Data were collected in visitor mode during 2006 March 26-28, using the
FORS2 spectrograph on UT1. The observed area covers the central region of
the Hydra I cluster, around NGC 3311, at $\alpha=$ 10$^h$36$^m$42.8$^s$,
$\delta$= -27$^d$31$^m$42$^s$ (J2000).
The FORS2 field of view (FOV) is $6'.8\times6'.8$ wide, corresponding
to $\sim$ 10000 kpc$^2$, and it is imaged onto a mosaic of two CCDs,
rebinned $2\times2$ in the readout.  We used two narrow band filters,
one centered at $\lambda=5045$ \AA\ and a second one at
$\lambda=5095$ \AA, both with a FWHM of $60$ \AA. In this way we are able 
to detect Ly$\alpha$ emission lines in the redshift range 3.12 $<$ z $<$ 3.21.
Spectra were obtained with the GRIS-600B grism, which has a
spectral resolution of 0.75 \AA/pixel at 5075 \AA. The MSIS mask is
made of $24\times21$ slits, each of them $0".8$ wide and $17".5$
long. Each slit is projected along the dispersion axis onto $\sim$ 40
rebinned pixels. The effective area imaged by the slits is $\sim$ 7056
arcsec$^2$, that is $\sim$ 4.5\% of the whole FORS2 FOV. In order to
cover the whole field, the MSIS mask was stepped 15 times on the sky to
fill the region between two adjacent slits. For each mask position 3
exposures, of $1200$ sec, were taken, ensuring a proper cosmic
ray subtraction.  The seeing during the three observing nights was in
the range from $0".6$ to $1".5$. We can detect emission line objects with a
flux completeness limit of $\sim$ 3.4 $\times$10$^{-18}$ erg cm$^{-2}$s$^{-1}$,
and their positions and radial velocities can be measured at the same time.
With this instrumental set up and total exposure time, monochromatic point
like emissions appear in the final images as unresolved sources with a width
of $\sim$ 6 pixels both in the spatial and in the wavelength direction.
Therefore, the FWHM spatial resolution is 1".1  and the FWMH spectral
resolution is $450$ km s$^{-1}$.

\section{Data reduction}
Data reduction was carried out following the procedure described in
\cite{ref:Arnaboldi07,ref:Ventimiglia08}. After bias subtraction, the images were
co-added and the continuum light from the two Hydra I galaxies was
subtracted, with an fmedian filtering using a window of $19\times35$
pixels.  The 2D stacked spectra of the emission sources were then
extracted, rectified, wavelength and flux calibrated. The total number
of emission sources detected is 86. On the basis of the flux
calibration the $1\sigma$ limit on the continuum is $\sim$
7$\times$10$^{-20}$ erg cm$^{-2}$ s$^{-1}$\AA$^{-1}$. Continuum fluxes
were calculated in the 60 \AA\ wavelength range covered by the filter
in which the emission line was detected.

In a first classification, all objects with unresolved emission lines and no
continuum were classified as PN candidates; the remainder as background galaxy
candidates. The PN candidate sample likely contains a fraction of unresolved
Ly$\alpha$ emitters without continuum which are not included in the present
study.

\section{Ly$\alpha$ candidates and their physical characteristics}
In the background galaxy sample 6 objects are classified as [OII]
emitters; they are characterized by an emission line with a clearly
visible continuum at all wavelengths.  The remaining 20 objects are
classified as candidate Ly$\alpha$ galaxies. They include sources with
resolved emission lines, either spatially or in wavelength or both,
without continuum, and sources with resolved/unresolved emission lines
and continuum only to the red side of the spectrum. This
classification is supported by spectroscopic follow up of similar
sample of background galaxies in previous PN surveys
\cite{ref:Freeman00,ref:Arnaboldi02,ref:Kudritzki00}, and by the fact
that the deep sample of $2.7<z< 3.7$ Ly$\alpha$ galaxies of
\cite{ref:Rauch08} contains few foreground [OII] emitting
galaxies, contrary to higher redshift samples \cite{ref:Martin08}.

\begin{figure}
 \includegraphics[width=6.5cm]{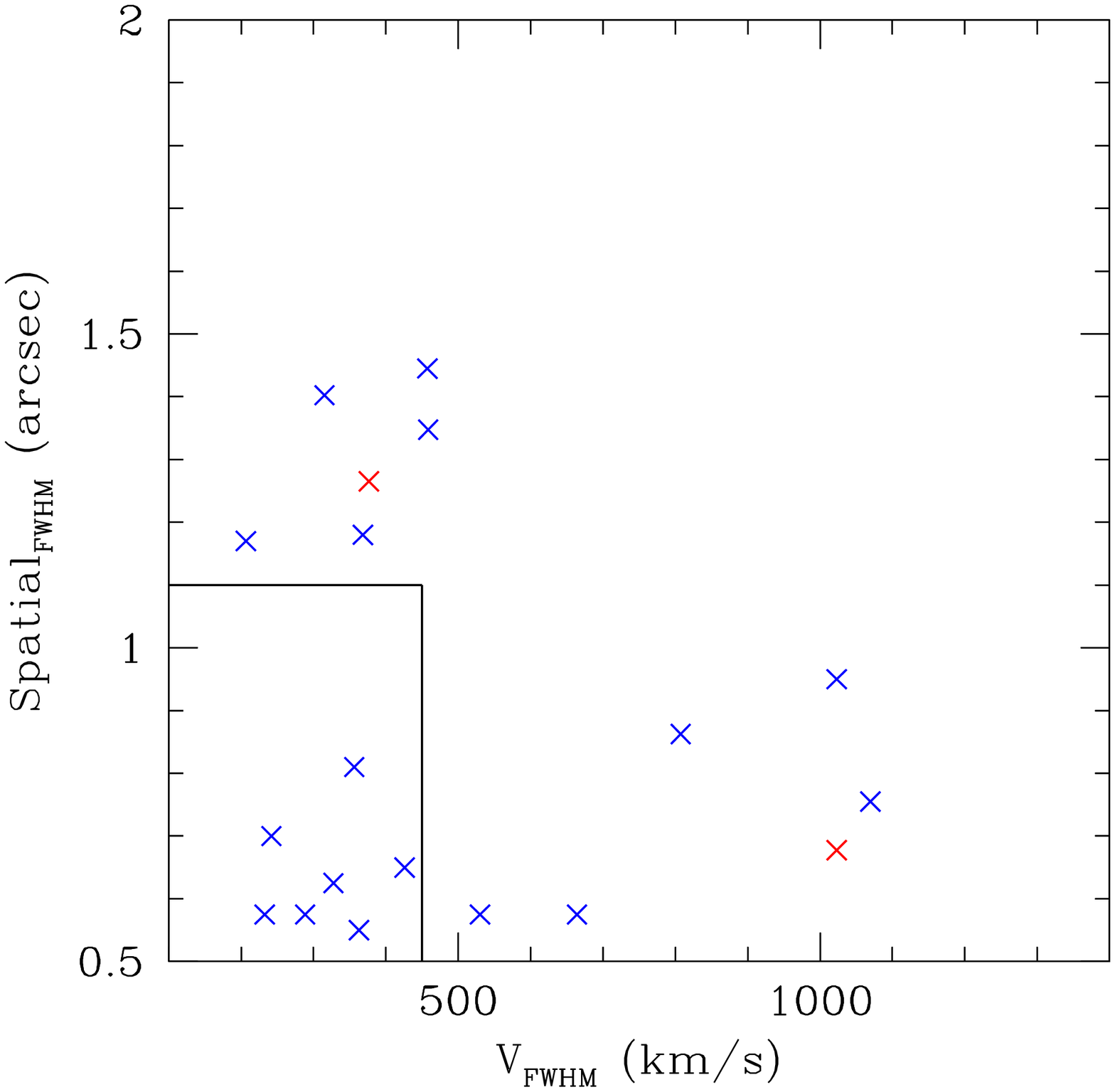}
 \includegraphics[width=6.5cm]{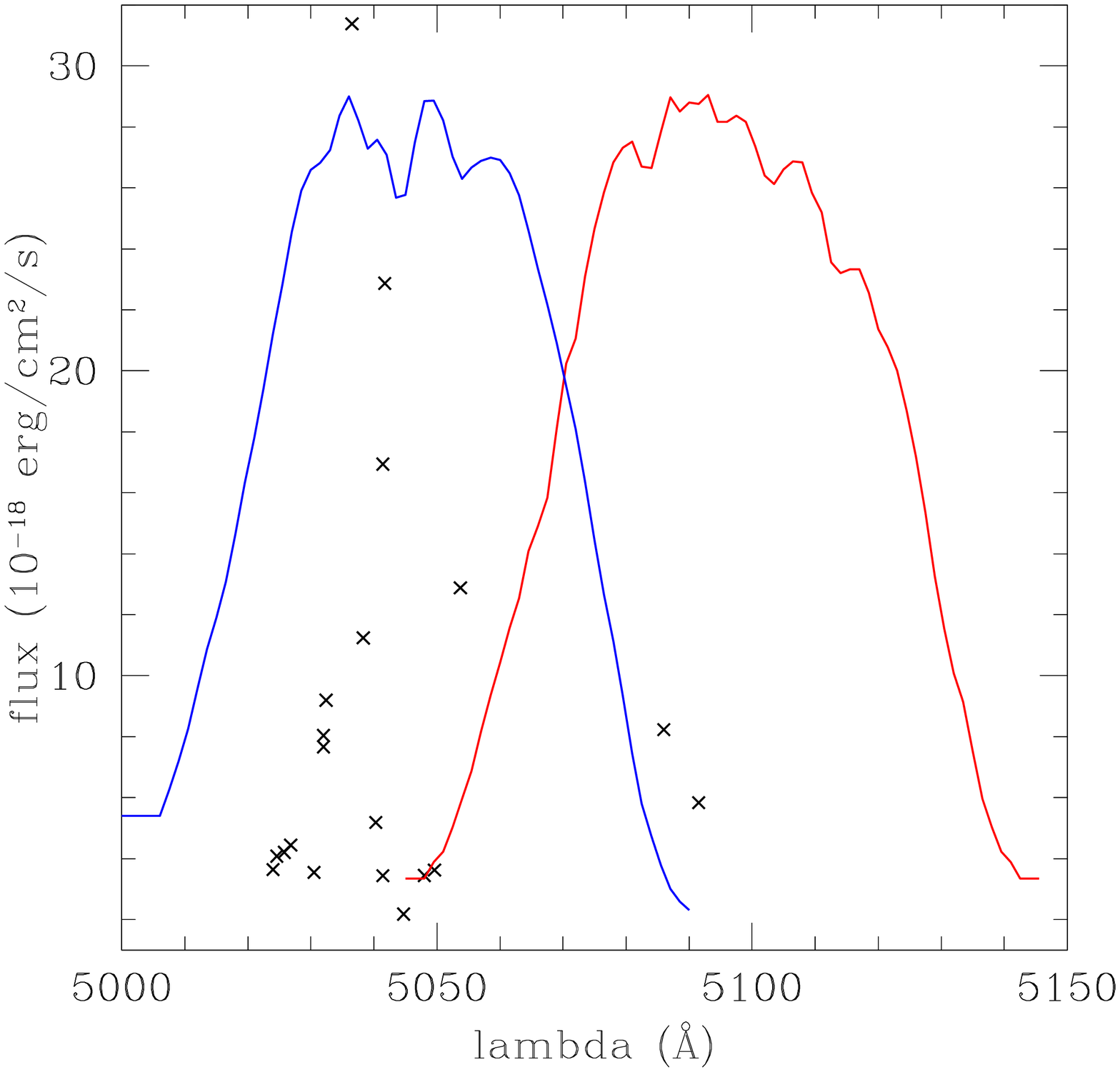}
 \includegraphics[width=6.5cm]{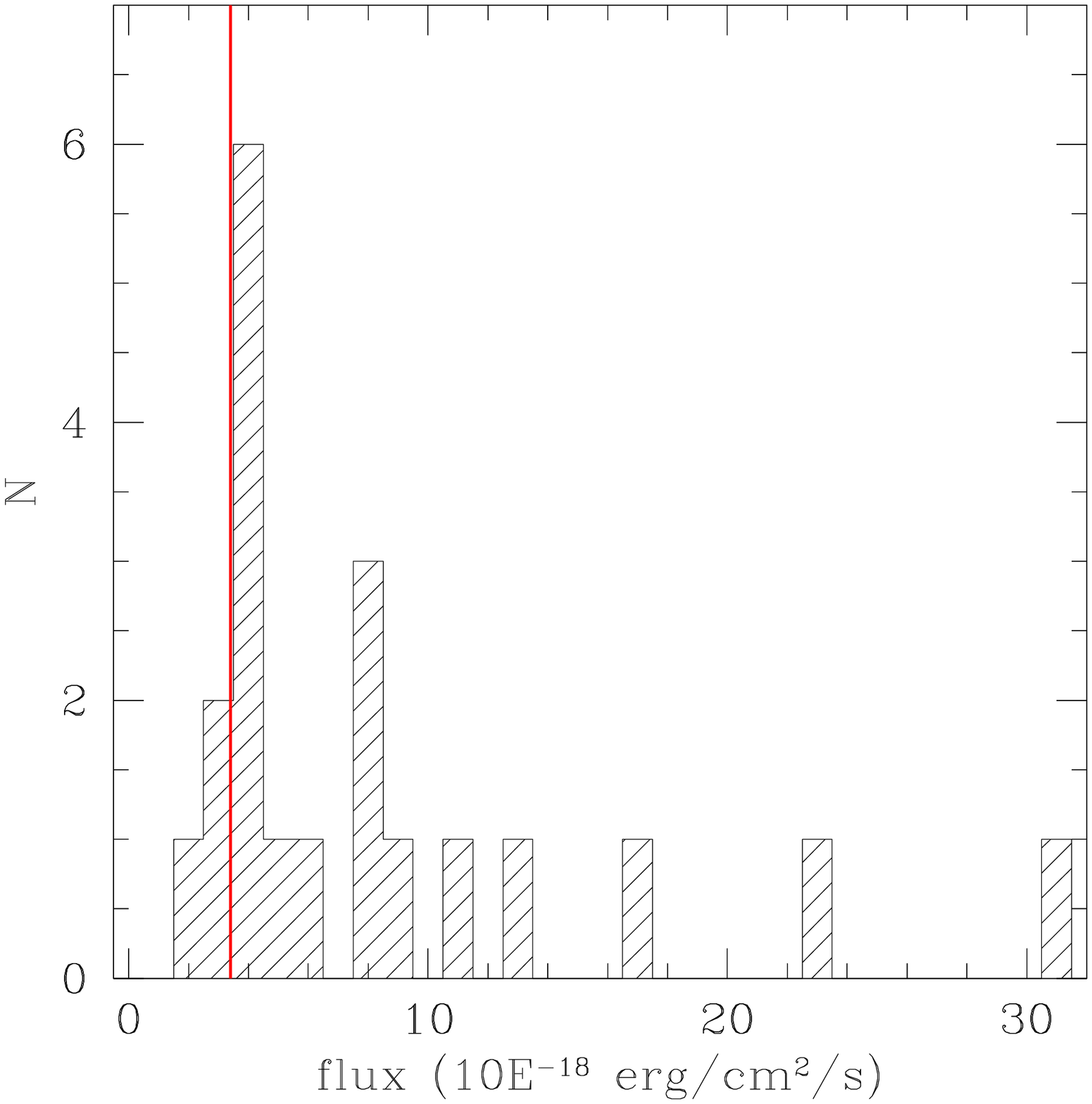}
 \includegraphics[width=6.5cm]{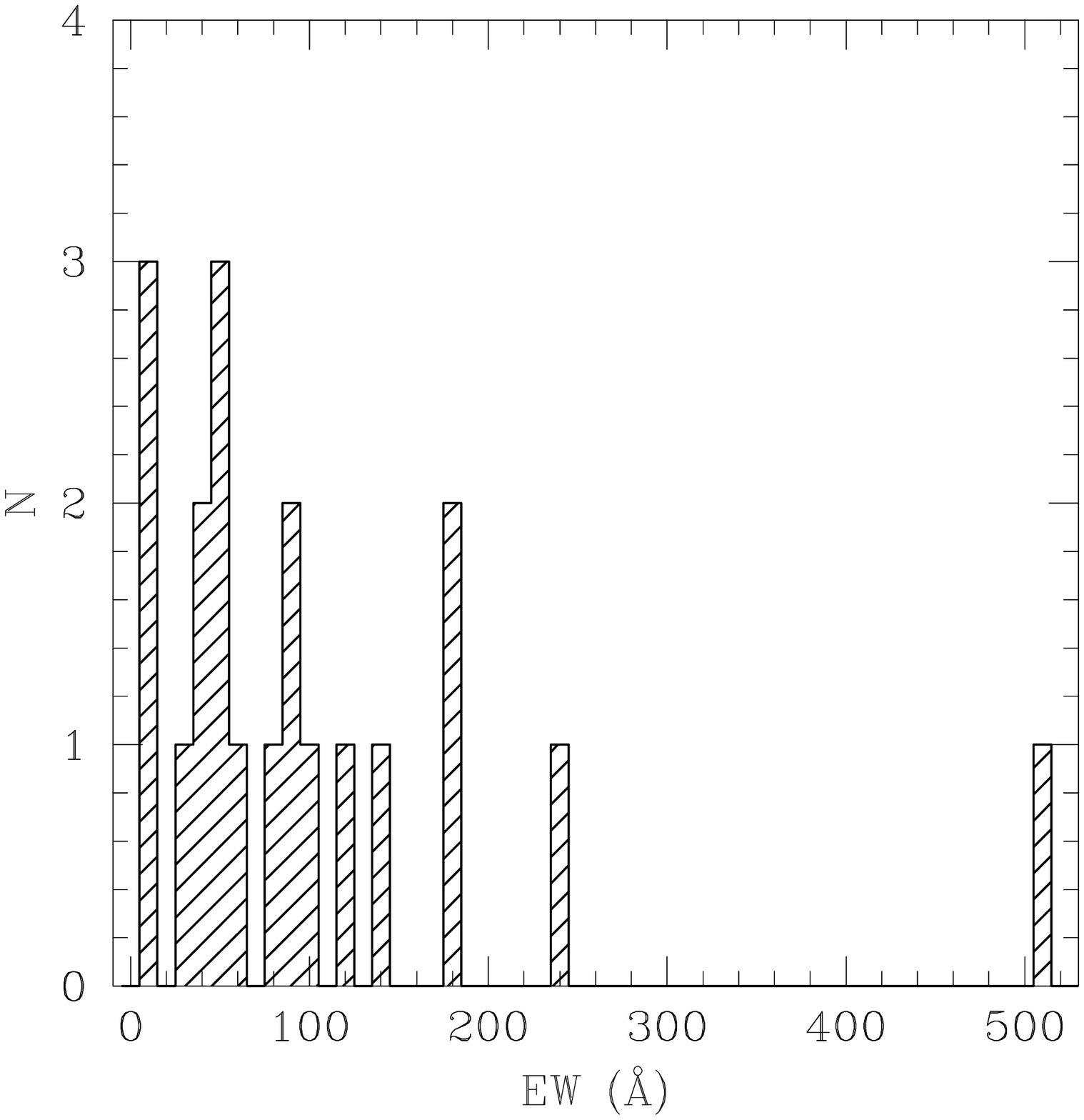}
 \caption{Physical characteristics of all emission line sources in the
   survey classified as Ly$\alpha$ candidate galaxies
   (LACs). \textit{Upper left panel}: Distribution of the Gaussian
   FWHM along the spatial direction vs.\ the Gaussian FWHM along the
   dispersion direction. One extended object in velocity could not be
   fitted by a Gaussian and is omitted.  In the rectangle on the lower
   left are LACs whose emission line is unresolved both spatially and
   in wavelength.  Blue crosses are objects detected in the wavelength
   range covered by the blue filter and red crosses are those detected
   in the wavelength range covered by the red filter. \textit{Upper
     right panel}: Flux vs.\ wavelength for the LACs (black crosses)
   detected in the MSIS field. The blue and the red lines show the
   normalized measured profiles of the blue and red filters,
   respectively.  \textit{Lower left panel}: Histogram of the LACs
   emission line flux distribution. The red vertical line defines the
   flux completeness limit.  \textit{Lower right panel}: Histogram of
   LACs EW distribution.}
 \label{FVFS}
 \end{figure}

The redshift distribution of the Ly$\alpha$ candidates (LACs) is shown
in Figure \ref{FVFS}.  The emission lines of all objects were fitted
by simple Gaussian profiles, both in the spatial and in the dispersion
direction.  For all the LACs, Fig.\ref{FVFS} (upper left panel) shows
the FWHM of the Gaussian profiles along the spatial direction vs.\ the
FWHM of the Gaussian profiles along the dispersion direction. About
65\% of the sample has an emission line which is spatially unresolved;
the rest is spatially resolved.  About 60\% of the sample is
unresolved in velocity and the rest is resolved.

 Based on this information we adopt a simple morphological
 classification of the candidates: unresolved objects both spatially
 and in wavelength (PS), resolved objects in both directions (EXT) and
 resolved objects only in the spatial or wavelength direction,
 respectively (EXTs, EXTv). A proper morphological classification is
 beyond the purpose of the current work. More detailed discussion
 about morphology and the related physics of LACs can be found in
 \cite{ref:Rauch08,ref:Verhamme08}.

 The fluxes were measured in apertures of $0".8\times2"\times755$ km/s
 centred around the emission lines and corrected for the filter
 response.  Fig.\ref{FVFS} shows the fluxes of the LACs as a function
 of wavelength (upper right panel), and their flux distribution (lower
 left panel). The latter is peaked at $\sim$ 3$\times$10$^{-18}$ erg
 cm$^{-2}$s$^{-1}$ and is truncated at faint fluxes, because our
 survey is flux limited. The completeness limit of the sample is
 $\sim$ 3.4$\times$10$^{-18}$ erg cm$^{-2}$s$^{-1}$, while the
 detection limit is $\sim$ 2$\times$10$^{-18}$ erg cm$^{-2}$s$^{-1}$.
 The most luminous objects have fluxes greater than
 20$\times$10$^{-18}$ erg cm$^{-2}$s$^{-1}$.

 The equivalent width (EW) distribution is also shown in
 Fig.\ref{FVFS} (lower right panel).  The continuum fluxes go from the
 limit on the 1$\sigma$ continuum value of 7$\times$10$^{-20}$ erg
 cm$^{-2}$s$^{-1}$ to $\sim$ 7$\times$10$^{-19}$ erg
 cm$^{-2}$s$^{-1}$. Most of the objects have an observed EW $<$ 200
 \AA.

A study of the luminosity function and a comparison of the number
densities with results from previous surveys at redshift $\sim$ 3.1 will be
the subject of a forthcoming paper.

\acknowledgments
G. Ventimiglia is supported by an ESO studentship.


\begin{thebibliography}{0}


\bibitem{ref:Ajiki03} \BY{Ajiki~M. et al.}
  \IN{AJ}{126}{2003}{091}


\bibitem{ref:Arnaboldi02} \BY{Arnaboldi~M. et al.}
  \IN{AJ}{123}{2002}{760}

\bibitem{ref:Arnaboldi07} \BY{Arnaboldi~M. et al.}
  \IN{PASJ}{59}{2007}{419}

\bibitem{ref:Ciardullo02} \BY{Ciardullo~R. et al.}
  \IN{ApJ}{566}{2002}{748}

\bibitem{ref:Freeman00} \BY{Freeman~K. et al.}
  \IN{ASP Conf. Series}{197}{2000}{389}

\bibitem{ref:Gerhard05} \BY{Gerhard~O. et al.}
  \IN{ApJ}{621}{2005}{L93}

\bibitem{ref:Gerhard07} \BY{Gerhard~O. et al.}
  \IN{A\&A}{468}{2007}{815}

\bibitem{ref:Gronwall07} \BY{Gronwall~C. et al.}
  \IN{ApJ}{667}{2007}{79}

\bibitem{ref:Giavalisco02a} \BY{Giavalisco~M.}
  \IN{ARA\&A}{40}{2002}{579}


\bibitem{ref:Hu98} \BY{Hu~E., Cowie~L. L. \atque McMahon~R. G.}
  \IN{ApJ}{502}{1998}{L99}

\bibitem{ref:Kudritzki00} \BY{Kudritzki~R.P. et al.}
 \IN{ApJ}{536}{2000}{536}

\bibitem{ref:Martin08} \BY{Martin~C.L., Sawicki~M., Dressler~A., McCarthy~P.}
  \IN{ApJ}{679}{2008}{942}

\bibitem{ref:Ouchi05} \BY{Ouchi~M. et al.}
  \IN{ApJ}{620}{2005}{L1}

\bibitem{ref:Ouchi08} \BY{Ouchi~M. et al.}
  \IN{ApJSS}{176}{2008}{301}


\bibitem{ref:Rauch08} \BY{Rauch~M. et al.}
  \IN{ApJ}{681}{2008}{856}

\bibitem{ref:Schaerer07} \BY{Schaerer~D.}
  \IN{Review}{706}{2007}{arXiv0706.0139S}

\bibitem{ref:Shapley03} \BY{Shapley~A. et al.}
  \IN{ApJ}{588}{2003}{65}

\bibitem{ref:Steidel96a} \BY{Steidel~C.C., Giavalisco~M., Dickinson~M., Adelberger~K.L.}
  \IN{AJ}{112}{1996}{352}

\bibitem{ref:Steidel96b} \BY{Steidel~C.C. et al.}
  \IN{ApJ}{462}{1996}{L17}

\bibitem{ref:Ventimiglia08} \BY{Ventimiglia~G. et al.}
  \IN{AN}{329}{2008}{1057}

\bibitem{ref:Verhamme08} \BY{Verhamme~A., Schaerer~D., Atek~H. \atque Tapken~C.}
  \IN{A\&A}{491}{2008}{89}


\end{thebibliography}
\end{document}